\documentclass[a4paper,twocolumn,showpacs,nofootinbib,floatfix,superscriptaddress,prc]{revtex4-1}
\usepackage{graphicx}
\usepackage{amsfonts}
\usepackage{amsmath}
\usepackage{hyperref}
\usepackage{float}

\begin{document}

\title{Fluid dynamical equations and transport coefficients of
relativistic gases with non-extensive statistics}
\author{T.\ S.\ Bir\'o$^{1}$ and E.\ Moln\'ar}

\affiliation{MTA KFKI, Research Institute for Particle and
Nuclear Physics, H-1525 Budapest, P.O.Box 49, Hungary}

\affiliation{Frankfurt Institute for Advanced Studies, Ruth-Moufang-Str.\ 1,
D-60438 Frankfurt am Main, Germany}

\begin{abstract}
We derive equations for fluid dynamics from a non-extensive Boltzmann
transport equation consistent with Tsallis' non-extensive entropy formula.
We evaluate transport coefficients employing the relaxation time
approximation and investigate non-extensive effects in leading order
dissipative phenomena at relativistic energies, like heat conductivity,
shear and bulk viscosity.
\end{abstract}

\maketitle

\section{Introduction}

The transport properties of matter at extreme high temperature and energy
density, well proved in high energy accelerator experiments at the 
Relativistic Heavy-Ion Collider (RHIC) and at the Large Hadron Collider (LHC),
raised some fundamental questions related to the smallest distance scale in physics. 
Theories, motivated by conformal field theory and higher dimensional dual gravity, 
in fact predicted a lower limit for the shear viscosity over entropy density ratio 
being around $1/4\pi$ in natural units \cite{AdSCFT}. 
Although this limitation can be overcome in more sophisticated, nonlinear 
dual gravity models \cite{Noronha:2009vz} or other approaches
\cite{Danielewicz:1984ww,Jakovac:2009nt,Jakovac:2009xn}, 
the actual calculation \cite{Csernai:2006zz,Xu:2007ns,NoronhaHostler:2008ju,Koide:2009sy,Chakraborty:2010fr,Bluhm:2010qf} 
and measurement of the viscosity and related properties of elementary matter remained 
in the focus of research interest.
 
The more, because the scientific evaluation of experimental signals, extracted from
analyses of particle spectra and correlations, can be interpreted in terms
of thermal concepts only if our knowledge about the material quality of a
(strongly) interacting quark-gluon plasma is well established \cite{Niemi:2011ix}. 
Besides the interpretation of temperature \cite{Biro:2009ki,Biro_book}, 
the whole hydrodynamical and statistical approach relies on our basic assumptions 
about the equation of state (EoS) and the dissipative capacities of this new stage of strongly
interacting matter \cite{Gyulassy:2004zy,Shuryak:2004cy}.

Modern statistical physics methods to the theoretical description of
high-energy phenomena include a consideration of non-extensive thermodynamics 
\cite{Tsallis,Tsallis_book,EPJA40_Tsallis,EPJA40_Beck,EPJA40_Kaniadakis,EPJA40_Kodama,EPJA40_Wilk,EPJA40_Alberico,EPJA40_Biro}. 
Besides giving a more general framework than Boltzmann did for the entropy formula, 
also the mesoscopic background including Langevin, Fokker-Planck or Boltzmann type 
equations can be extended to involve deviations from the classical nineteenth century
picture \cite{Biro:2004qg,Biro:2005uv,Kaniadakis_2001,Walton:1999dy,Sherman:2002vi}. 
Most of these generalizations depend on a single parameter, denoted by $q$, which
also influences the canonical equilibrium state leading from an exponential
Gibbs distribution in energy to a power-law tailed one. 
These are observable in experimental transverse momentum spectra and
multiplicity fluctuations \cite{Beck_2003,Biro:2008km,Wilk:1999dr,Wilk:2006dv}, 
therefore the important question arises, how are the kinetic and hydrodynamical 
approaches affected by the value of this parameter. 
Are there qualitative or only quantitative, minor deviations to be expected when $q\neq 1$? 
Fits to the Relativistic Heavy-Ion Collider (RHIC) and the Large Hadron Collider (LHC)
spectra make a hadronic matter value of $q_{H}\approx 1.08..1.2$
\cite{Tang:2008ud,Shao:2009mu,Sikler:HCBM} and a quark matter value 
of $q_{Q}\approx 1.22$ \cite{Biro:2008er} probable.

These fits point out a deviation from the standard Boltzmann-Gibbs (BG)
statistics for $q=1$. The degree of non-equilibrium in a system is defined
relative to a fixed local $q$-equilibrium. Assuming $q\neq 1$ means that the
local equilibrium differs from the standard BG equilibrium. On the other
hand it may be regarded as non-equilibrium compared to the classical $q=1$,
if one considers a time evolution of this parameter. However, we do not
consider this possibility in the present paper.

We investigate a general non-equilibrium system which - for any given $q$ -
dissipates energy and produces entropy. This we call $q$-non-equilibrium.
Similarly to classical BG systems this $q$-non-equilibrium is related to the
response of the system to gradients of different thermodynamical intensives.
This $q$-non-equilibrium state does not relax to a standard BG equilibrium,
but to a local $q$-equilibrium.

In this paper we derive the relativistic fluid-dynamical equations of motion
from a non-extensive relativistic Boltzmann transport equation (NEBE) and
calculate the transport coefficients employing the relaxation time
approximation for the collision integral. 
Such calculations were done in the non-relativistic limit \cite{Trans_coeffs} 
with a non-extensive Boltzmann equation \cite{Lima_2001,Lima_2005} slightly 
different from ours. 
The $q$-generalized first order relativistic Navier-Stokes-Fourier equations of
relativistic dissipative fluid dynamics are also derived within this framework. 
We use the metric of flat space-time $g^{\mu \nu }\equiv $diag$%
(1,-1,-1,-1)$ and natural units throughout this work, $\hbar =k_{B}=c=1$.

\section{Non-Extensive Boltzmann Equation and relativistic hydrodynamics}

We follow Lavagno \cite{Lavagno_2002} and recall the $q$-modified
non-extensive Boltzmann equation (NEBE) compatible with Tsallis' suggestion
for a generalized non-extensive entropy formula. 
Extensions of the classical Boltzmann equation, like the one we adopt here 
from \cite{Lavagno_2002}, are put forward without any derivation from microscopic quantum field dynamics, 
with the purpose to study classical statistical effects due to correlations possibly 
induced by finite system size compared to the characteristic interaction range \cite{Tsallis_book}.

For the sake of simplicity we neglect external field effects which simplifies 
the Vlasov operator in the following general Boltzmann-type equation: 
\begin{equation}
k^{\mu }\partial _{\mu }\tilde{f}_{k}=C\left[ f\right] ,
\end{equation}%
where $f_{k}\equiv f\left( x^{\mu },k^{\mu }\right)$ is the single particle
phase-space distribution function and $\tilde{f}_{k}=\left( f_{k}\right)^{q}$, 
with $q\in \left( 0,2\right) $. $C\left[ f\right] $ is the collision integral. 
The space-time coordinates are $x^{\mu }=(t,\mathbf{x})$ and the
particle four-momentum $k^{\mu }=(k^{0},\mathbf{k})$ is normalized to the
mass $m=\sqrt{k^{\mu }k_{\mu }}$. The NEBE is a closed equation assuming
that the number of binary collisions around space-time coordinates $x^{\mu }$
is proportional to an expression $H_{q}\left[ f_{k},f_{k^{\prime }}\right] $, 
where the explicit form of the collision integral involving binary
collision with incoming momenta $p^{\mu }$, $p^{\prime \mu }$ and outgoing
momenta $k^{\mu }$, $k^{\prime \mu }$ is given as, 
\begin{align}
C\left[ f\right] & =\frac{1}{2}\int dK^{\prime }dPdP^{\prime }  \notag \\
& \times W_{kk\prime \rightarrow pp\prime }\left( H_{q}\left[
f_{p},f_{p^{\prime }}\right] - H_{q}\left[ f_{k},f_{k^{\prime }}\right]
\right) .
\end{align}
Here the factor $1/2$ takes into account the symmetry of the initial or
final momentum configuration in case of identical particles if we
interchange the labels i.e., $\ \left( k^{\mu },k^{\prime \mu }\right)
\leftrightarrow $ $\left( k^{\prime \mu },k^{\mu }\right) $ or $\left(
p^{\mu },p^{\prime \mu }\right) \leftrightarrow $ $\left( p^{\prime \mu
},p^{\mu }\right) $. Furthermore, $dK\equiv gd^{3}\mathbf{k/}\left( (2\pi
\hbar )^{3}k^{0}\right) $ is the Lorentz-invariant momentum-space volume,
with $g$ being the number of internal degrees of freedom, e.g. spin
degeneracy, while $W_{kk\prime \rightarrow pp\prime }$ is the
Lorentz-invariant transition rate. The transition rate is symmetric with
respect to the sequence of final states $W_{kk\prime \rightarrow pp\prime
}=W_{kk\prime \rightarrow p\prime p}$, as well as it is symmetric for time
reversed processes $W_{kk\prime \rightarrow pp\prime }=W_{pp\prime
\rightarrow kk\prime }$. 
In this way it fulfils the detailed balance property. 
The collision integral is positive for any $f\geq 0$ and it does 
not change sign under time reversal. 
The $q$-generalized version of the so-called Stosszahlansatz, 
the assumption of molecular chaos, 
\begin{equation}
H_{q}\left[ f_{k},f_{k^{\prime }}\right] \equiv \exp _{q}\left[ \ln
_{q}\left( f_{k}\right) +\ln _{q}\left( f_{k^{\prime }}\right) \right] ,
\end{equation}%
uses the $q$-deformed exponential and logarithm functions, 
\begin{eqnarray}
\exp _{q}\left( x\right) &\equiv &\left[ 1+\left( 1-q\right) x\right]
^{1/\left( 1-q\right) }, \\
\ln _{q}\left( x\right) &\equiv &\frac{x^{1-q}-1}{1-q}.
\end{eqnarray}%
For $q\rightarrow 1$ the above expressions are the standard exponential and
natural logarithm functions. 
Here we stress that for $q=1$ the detailed balance solution of the above
equations obey the classical Boltzmann-Gibbs statistics, not the
usual quantum statistics.

The entropy four-current has to be properly defined as 
\begin{equation}
S^{\mu }\equiv -\int dK\ k^{\mu }\left[ \tilde{f}_{k}\ln _{q}\left(
f_{k}\right) -f_{k}\right] ,  \label{kinetic_S_mu}
\end{equation}%
demanding positive entropy production according to the second law of
thermodynamics: 
\begin{equation}
\partial _{\mu }S^{\mu }\equiv -\int dK\ln _{q}\left( f_{k}\right) \left[
k^{\mu }\partial _{\mu }\tilde{f}_{k}\right] \geq 0.
\label{kinetic_entropy_production}
\end{equation}%
Note that Lavagno defined the entropy four-current as $S_{L}^{\mu }\equiv
-\int dK\ k^{\mu }\tilde{f}_{k}\left[ \ln _{q}\left( f_{k}\right) -1\right]$, 
while the current form was introduced, correcting Lavagno's definition, by
Osada and Wilk \cite{Osada:2008sw,Osada:2008hs,Osada:2008cn,Osada:2009cc}. 
We shall return to this point later.

In order to elucidate the relation to a fluid dynamical description, let us
define now the following momentum-space integrals of the NEBE, 
\begin{eqnarray}
\partial _{\mu }N^{\mu } &\equiv &\int dK\ k^{\mu }\partial _{\mu }\tilde{f}%
_{k}=\int dKC\left[ f\right] ,  \label{kinetic_N_mu_cons} \\
\partial _{\mu }T^{\mu \nu } &\equiv &\int dK\ k^{\nu }k^{\mu }\partial
_{\mu }\tilde{f}_{k}=\int dKk^{\mu }C\left[ f\right] ,
\label{kinetic_T_mu_nu_cons}
\end{eqnarray}%
where we introduced the $q$-modified particle four-current and symmetric
energy-momentum tensor, 
\begin{eqnarray}
N^{\mu } &\equiv &\int dK\ k^{\mu }\tilde{f}_{k},  \label{kinetic_N_mu} \\
T^{\mu \nu } &\equiv &\int dK\ k^{\mu }k^{\nu }\tilde{f}_{k}.
\label{kinetic_T_mu_nu}
\end{eqnarray}%
It is straightforward to show that the left hand side of Eqs. (\ref%
{kinetic_N_mu_cons}) and (\ref{kinetic_T_mu_nu_cons}) vanishes when the
particle number, the energy and momentum in individual collisions are
conserved, that is, if $p^{\mu }+p^{\prime \mu }=k^{\mu }+k^{\prime \mu }$
holds. This property is expressed by specifying the so-called collision
invariants by using, $\psi =\alpha +\beta _{\mu }k^{\mu }$, such that 
\begin{align}
\int dK\alpha C\left[ f\right] & =0, \\
\int dK\beta _{\mu }k^{\mu }C\left[ f\right] & = 0.
\end{align}%
It implies the macroscopic conservation laws of particle four-current and
energy-momentum tensor for any solution of the NEBE, 
\begin{eqnarray}
\partial _{\mu }N^{\mu } &=&0,  \label{N_mu_cons} \\
\partial _{\mu }T^{\mu \nu } &=&0,  \label{T_mu_nu_cons}
\end{eqnarray}%
besides a non-negative entropy production from Eq. ($\ref%
{kinetic_entropy_production}$).

The entropy production vanishes in local $q$-equilibrium and it leads to
the collision invariant $\psi =\ln _{q}\left( f_{0k}\right)$. The
distribution function which satisfies $\partial _{\mu }S^{\mu }\left(
f_{0k}\right) =0$ is the canonical equilibrium one. This result is
equivalent with the requirement that $\ln _{q}\left( f_{0k}\right) =\alpha
_{0}+\beta _{0}^{\mu }k_{\mu }$, whence the $q$-equilibrium distribution is
given by 
\begin{equation}
f_{0k}=\exp_{q}\left( \alpha _{0}-\beta _{0}k^{\mu }u_{\mu }\right) .
\label{q_equilibrium}
\end{equation}%
Here $\alpha =\alpha _{0}$ and $\beta ^{\mu }=\beta _{0}u^{\mu }$ are the
collisional invariants given previously, and $u^{\mu }$ is a four-vector
normalized to one, $u^{\mu }u_{\mu }=1$. These quantities will be identified
with the inverse temperature $\beta _{0}=1/T$, the chemical potential over
temperature, $\alpha _{0}=\mu /T$, and the fluid dynamical four-velocity of
matter $u^{\mu }$. The formula in Eq. (\ref{q_equilibrium}) reduces to the
well known J\"{u}ttner distribution or relativistic Maxwell-Boltzmann
distribution for $q=1$, that is $f_{J}=\exp \left( \alpha _{0}-\beta
_{0}k^{\mu }u_{\mu }\right) $. Therefore it is clear that in local $q$%
-equilibrium there are only five fields which completely characterize the
system for any given $q$.

For the purpose of a quick reference let us introduce $\Delta ^{\mu \nu
}\equiv g^{\mu \nu }-u^{\mu }u^{\nu }$, used to project an arbitrary
four-vector into another four-vector orthogonal to $u^{\mu }$. Any
four-vector, $A^{\mu }$, and in particular the four-momenta of particles can
be decomposed into two parts using an arbitrary fluid dynamic flow velocity, 
$u^{\mu }$: $k^{\mu }=E_{k}u^{\mu }+k^{\left\langle \mu \right\rangle }$,
where $E_{k}=k^{\mu }u_{\mu }$ is the Local Rest Frame (LRF) energy of the
particle and $k^{\left\langle \mu \right\rangle }=\Delta ^{\mu \nu }k_{\nu }$
contains the LRF momenta \cite{Anderson_1}. 
The LRF or co-moving frame is defined with $u_{LRF}^{\mu }=\left( 1,0,0,0\right)$.
Moreover, the space-time gradient or four-divergence, $\partial _{\mu
}\equiv \partial /\partial x^{\mu }=u_{\mu }d/d\tau +\nabla _{\mu }$, should
also be decomposed into parts parallel and orthogonal to the flow. Here, the
co-moving or proper time-derivative, $d/d\tau \equiv u^{\mu }\partial _{\mu
} $, is also denoted by an over-dot, $\dot{A}^{\mu }\equiv dA^{\mu }/d\tau $%
, and $\nabla _{\mu }\equiv \Delta _{\mu }^{\nu }\partial _{\nu }$ denotes
the gradient. For second-rank tensors the orthogonal and traceless
projection is defined as $A^{\left\langle \mu \nu \right\rangle }=\Delta ^{\mu \nu \alpha
\beta }A_{\alpha \beta }$, where $\Delta ^{\mu \nu \alpha \beta }\equiv 
\frac{1}{2}\left( \Delta ^{\mu \alpha }\Delta ^{\beta \nu }+\Delta ^{\nu
\alpha }\Delta ^{\beta \mu }\right) -\frac{1}{3}\Delta ^{\mu \nu }\Delta
^{\alpha \beta }$.

Now, using the local $q$-equilibrium distribution function we are prepared
to introduce the following generalized thermodynamic integrals:
\begin{align}
\mathcal{I}_{q\left( i,j\right) }& \equiv \frac{1}{\left( 2j+1\right) !!}%
\int dK\left( E_{k}\right) ^{i-2j}\left( \Delta ^{\mu \nu }k_{\mu }k_{\nu
}\right) ^{j}f_{0k},  \\
\mathcal{J}_{q\left( i,j\right) }& \equiv \frac{1}{\left( 2j+1\right) !!}%
\int dK\left( E_{k}\right) ^{i-2j}\left( \Delta ^{\mu \nu }k_{\mu }k_{\nu
}\right) ^{j}\left( f_{0k}\right) ^{q},  \label{J_i_j} \\
\mathcal{K}_{q\left( i,j\right) }& \equiv \frac{q}{\left( 2j+1\right) !!}%
\int dK\left( E_{k}\right) ^{i-2j}\left( \Delta ^{\mu \nu }k_{\mu }k_{\nu
}\right) ^{j}\left( f_{0k}\right) ^{2q-1},  \label{K_i_j}
\end{align}%
where $i,j\geq 0$ are natural numbers and $\left( 2j+1\right) !!=\left(
2j+1\right) !/\left( j!2^{j}\right) $ denotes the double factorial. At a
given fixed temperature $\left( f_{0k}\right) ^{q}=\left( \partial
f_{0k}/\partial \alpha _{0}\right)|_{\beta _{0}}$ and $q\left( f_{0k}\right)
^{2q-1}=\left( \partial \left( f_{0k}\right) ^{q}/\partial \alpha
_{0}\right)| _{\beta _{0}}$, whence it follows that 
\begin{equation}
\mathcal{J}_{q\left( i,j\right) }=\left( \frac{\partial \mathcal{I}_{q\left(
i,j\right) }}{\partial \alpha _{0}}\right) _{\beta _{0}},\ \mathcal{K}%
_{q\left( i,j\right) }=\left( \frac{\partial \mathcal{J}_{q\left( i,j\right)
}}{\partial \alpha _{0}}\right) _{\beta _{0}}.
\end{equation}

The above integrals for $q<1$ run over a finite range in energy, $%
E_{k}\leq T/(1-q)$, and are always convergent. For $q>1$ the integrands have
power-law tails satisfying a convergence criterion: $(i+2)<1/(q-1)$ is
required for the $\mathcal{I}_{q}(i,j)$ integrals to be finite.
Correspondingly the maximal index $i$ is one higher for the $\mathcal{J}_{q}$
and two higher for the $\mathcal{K}_{q}$ integrals. Finally, for $q=1$ the
familiar J\"{u}ttner distribution makes all integrals convergent by its
exponential tail for any finite indices. In this case 
$\mathcal{I}_{q=1}(i,j)=\mathcal{J}_{q=1}(i,j)=\mathcal{K}_{q=1}(i,j)=I(i,j)$ 
we retrieve the familiar relativistic thermodynamic integrals, 
see Refs. \cite{deGroot_book,Cercignani_book,Muronga:2006zx}.
Note that in the paper by Lavagno 
\cite{Lavagno_2002,Lavagno_2009}, the $q $-modified Bessel function of second kind was
introduced, $K_{n}(q,z)$, where $z=m\beta _{0}$, which in our notation
corresponds to $\mathcal{J}_{q\left( i,j\right) }$ in the LRF.

It is also straightforward to show by partial integration that the following
recursive relations hold in equilibrium: 
\begin{eqnarray}
\mathcal{J}_{q\left( i,j\right) } &=&-\frac{1}{\beta _{0}}\mathcal{I}%
_{q\left( i-1,j-1\right) }+\frac{i-2j}{\beta _{0}}\mathcal{I}_{q\left(
i-1,j\right) },  \label{Jq_nk_rec} \\
\mathcal{K}_{q\left( i,j\right) } &=&-\frac{1}{\beta _{0}}\mathcal{J}%
_{q\left( i-1,j-1\right) }+\frac{i-2j}{\beta _{0}}\mathcal{J}_{q\left(
i-1,j\right) }. \label{Kq_nk_rec}
\end{eqnarray}%
Furthermore, one can also show that
\begin{eqnarray}
\mathcal{I}_{q\left( i+2,j\right) } &=&m^{2}\mathcal{I}_{q\left( i,j\right)
}-\left( 2j+3\right) \mathcal{I}_{q\left( i+2,j+1\right) }, \\
\mathcal{J}_{q\left( i+2,j\right) } &=&m^{2}\mathcal{J}_{q\left( i,j\right)
}-\left( 2j+3\right) \mathcal{J}_{q\left( i+2,j+1\right) }, \\
\mathcal{K}_{q\left( i+2,j\right) } &=&m^{2}\mathcal{K}_{q\left( i,j\right)
}-\left( 2j+3\right) \mathcal{K}_{q\left( i+2,j+1\right) },
\end{eqnarray}%
as well as%
\begin{eqnarray}
\dot{\mathcal{I}}_{q(i,j)} &\equiv &\frac{\partial \mathcal{I}_{q\left(
i,j\right) }}{\partial \alpha _{0}}\dot{\alpha}_{0}+\frac{\partial \mathcal{I%
}_{q\left( i,j\right) }}{\partial \beta _{0}}\dot{\beta}_{0}  \notag \\
&=&\mathcal{J}_{q\left( i,j\right) }\dot{\alpha}_{0}-\mathcal{J}_{q\left(
i+1,j\right) }\dot{\beta}_{0},
\end{eqnarray}%
and similarly
\begin{eqnarray}
\dot{\mathcal{J}}_{q(i,j)} &\equiv &\frac{\partial \mathcal{J}_{q\left(
i,j\right) }}{\partial \alpha _{0}}\dot{\alpha}_{0}+\frac{\partial \mathcal{J%
}_{q\left( i,j\right) }}{\partial \beta _{0}}\dot{\beta}_{0} \notag \\
&=&\mathcal{K}_{q\left( i,j\right) }\dot{\alpha}_{0}-\mathcal{K}_{q\left(
i+1,j\right) }\dot{\beta}_{0}.  \label{d_J_nk}
\end{eqnarray}

Let us recall the definition of the $q$-modified entropy four-current and
calculate the equilibrium entropy density, $s_{0}=S_{0}^{\mu }u_{\mu }$ with 
$S_{0}^{\mu }=S^{\mu }\left( f_{0k}\right) $ for the $q$-equilibrium
distribution function from Eq. (\ref{q_equilibrium}). Making use of the
previously introduced $q$-generalized thermodynamic integrals we obtain 
\begin{align}
s_{0}& \equiv -\alpha _{0}\int dKE_{k}\tilde{f}_{0k}+\beta _{0}\int
dKE_{k}^{2}\tilde{f}_{0k}+\int dKE_{k}f_{0k}  \notag \\
& =-\alpha _{0}\mathcal{J}_{q\left( 1,0\right) }+\beta _{0}\mathcal{J}%
_{q\left( 2,0\right) }+\mathcal{I}_{q\left( 1,0\right) }.  \label{EQ27}
\end{align}%
Identifying the momentum integrals, or shortly "the moments", as the
particle density $n_{0}=\mathcal{J}_{q\left( 1,0\right) }$, energy density $%
e_{0}=\mathcal{J}_{q\left( 2,0\right) }$, and pressure $p_{0}=-\mathcal{J}%
_{q\left( 2,1\right) }$. Here we have used Eq. (\ref{Jq_nk_rec}) to obtain $%
\mathcal{J}_{q\left( 2,1\right) }=-\beta _{0}^{-1}\mathcal{I}_{q\left(
1,0\right) }$ which translates into the familiar ideal gas EoS, 
$p_{0}=\beta _{0}^{-1}\mathcal{I}_{q\left( 1,0\right) }$. This is the
reason why the definition of the entropy four-current has to be given in the
specific form of Eq. (\ref{kinetic_S_mu}). Hence the fundamental
thermodynamic equation follows directly from Eq. (\ref{EQ27}): 
\begin{equation}
s_{0}=-\alpha _{0}n_{0}+\beta _{0}\left( e_{0}+p_{0}\right) .
\label{entropy}
\end{equation}%
Note that $\beta _{0}p_{0}\neq n_{0}$, because $n_{0}=\mathcal{J}_{q\left( 1,0\right)
}\neq \mathcal{I}_{q\left( 1,0\right) }$. However, if the number of
particles is conserved, the integrals of $\mathcal{I}_{q}$ and $\mathcal{J}_{q}$ 
should lead the same number of particles per unit volume, although the
normalization of the distribution functions differ.

The definition of entropy leads to the fundamental thermodynamic relation and
the well known Gibbs-Duhem relations, 
\begin{eqnarray}
ds_{0} &=&-\alpha _{0}dn_{0}+\beta _{0}de_{0}, \\
dp_{0} &=&\frac{n_{0}}{\beta _{0}}d\alpha _{0}-\frac{\left(
e_{0}+p_{0}\right) }{\beta _{0}}d\beta _{0}.
\end{eqnarray}
The last equation follows from Eqs. (\ref{Kq_nk_rec},\ref{d_J_nk}).
Note that, the above thermostatic relations resemble those of 
classical BG thermodynamics. 
Kinetic theory does not specify the zeroth law of thermodynamics, therefore, whether 
the thermodynamic temperature is different from the temperature 
in kinetic theory is an open question \cite{BiroVanPRE2011}.
However, choosing a single parameter for the temperature, the equations 
of $q$-fluid dynamics become formally identical with that of classical 
fluid dynamics for $q=1$.

Before going further we discuss the fluid dynamical equations in $q$%
-equilibrium. The particle four-current and energy-momentum tensor are
calculated from Eqs. (\ref{kinetic_N_mu},\ref{kinetic_T_mu_nu}) using the
local $q$-equilibrium distribution function from Eq. (\ref{q_equilibrium}),
hence 
\begin{eqnarray}
N_{0}^{\mu } &=&n_{0}u^{\mu },  \label{N_mu_equilibrium} \\
T_{0}^{\mu \nu } &=&e_{0}u^{\mu }u^{\nu }-p_{0}\Delta ^{\mu \nu }.
\label{T_mu_nu_equilibrium}
\end{eqnarray}%
These decompositions are formally identical with that of a perfect fluid,
hence some authors called it a perfect q-fluid \cite{Osada:2008cn,Osada:2009cc}.
Moreover, the conservation laws from Eqs. (\ref{N_mu_cons}, \ref%
{T_mu_nu_cons}) lead, at least formally, to the well known Euler equations
of perfect fluid dynamics, 
\begin{eqnarray}
\partial _{\mu }N_{0}^{\mu } &=&0, \\
\partial _{\mu }T_{0}^{\mu \nu } &=&0,
\end{eqnarray}%
making a closed system of equations by supplementing with an EoS. These
equations imply a vanishing local entropy production in $q$-local
equilibrium, $\partial _{\mu }S_{0}^{\mu }=0$ for smooth initial conditions,
i.e. without discontinuities, and EoS's without first order phase
transitions. Note that in general the EoS is not restricted to be an ideal gas EoS. In
our case the EoS is given as, $p_{0}=\frac{1}{3}\left( e_{0}-m^{2}\mathcal{J}%
_{q\left( 0,0\right) }\right) $, which in the massless limit returns the
familiar $e_{0}=3p_{0}$ and hence the speed of sound is, $c_{s}=\sqrt{1/3}$,
irrespective of the $q$-parameter \cite{Lavagno_2002}.

On this account the general formula can be derived rewriting the above
equations with the help of the thermodynamic integrals and relations, $%
d\alpha _{0}=\left( de_{0}+\mathcal{K}_{q\left( 3,0\right) }d\beta
_{0}\right) /\mathcal{K}_{q\left( 2,0\right) }$ and $d\beta _{0}=-\left(
de_{0}-\mathcal{K}_{q\left( 2,0\right) }d\alpha _{0}\right) /\mathcal{K}_{q\left( 3,0\right) }$. 
After some algebra we get that the speed of sound
squared $c_{s}^{2}=\left( \frac{\partial p_{0}}{\partial e_{0}}\right) $ at
fixed entropy per particle, $s_0/n_0$, is given as,%
\begin{align}
c_{s}^{2}& =\frac{1}{3}+\frac{m^{2}}{3}\frac{\mathcal{D}_{q\left( 1,0\right)
}}{\mathcal{D}_{q\left( 2,0\right) }}  \notag \\
& -\frac{m^{2}}{3}\frac{\left( \mathcal{K}_{q\left( 3,0\right) }\mathcal{K}%
_{q\left( 0,0\right) }-\mathcal{K}_{q\left( 2,0\right) }\mathcal{K}_{q\left(
1,0\right) }\right) }{h_{0}\mathcal{D}_{q\left( 2,0\right) }},  \label{cs_2}
\end{align}%
where $h_{0}\equiv \mathcal{K}_{q\left( 3,1\right) }/\mathcal{K}_{q\left(
2,1\right) }=\left( e_{0}+p_{0}\right) /n_{0}$ is the enthalpy per particle
and $\mathcal{D}_{q\left( 1,0\right) }=\mathcal{K}_{q\left( 2,0\right) }%
\mathcal{K}_{q\left( 0,0\right) }-\mathcal{K}_{q\left( 1,0\right) }^{2}$, $%
\mathcal{D}_{q\left( 2,0\right) }=\mathcal{K}_{q\left( 3,0\right) }\mathcal{K%
}_{q\left( 1,0\right) }-\mathcal{K}_{q\left( 2,0\right) }^{2}$.


\section{Beyond q-equilibrium}

If a system is out-of-equilibrium the distribution function is different
from the local$\ q$-equilibrium distribution, $\tilde{f}_{k}\neq \tilde{f}%
_{0k}$. This will lead to additional terms in the thermodynamic quantities
and increase the entropy until the system relaxes to local $q$%
-equilibrium according to the microscopic dynamics described by a NEBE. Once
local $q$-equilibrium is reached these additional dissipative quantities vanish.

Making use of the previously introduced notations, the macroscopic fields,
such as the particle four-current and energy-momentum tensor, can be
decomposed in a general frame the following way, 
\begin{align}
N^{\mu }& \equiv u^{\mu }\int dKE_{k}\tilde{f}_{k}+\int dKk^{\left\langle
\mu \right\rangle }\tilde{f}_{k}, \\
T^{\mu \nu }& \equiv u^{\mu }u^{\nu }\int dKE_{k}^{2}\tilde{f}_{k}+\frac{1}{3%
}\Delta ^{\mu \nu }\int dK\left( \Delta ^{\alpha \beta }k_{\alpha }k_{\beta
}\right) \tilde{f}_{k}  \notag \\
& +2u^{\left( \mu \right. }\int dKE_{k}k^{\left. \left\langle \nu
\right\rangle \right) }\tilde{f}_{k}+\int dKk^{\left\langle \mu \right.
}k^{\left. \nu \right\rangle }\tilde{f}_{k},
\end{align}%
where the round brackets around the greek indices denote symmetrization: $%
A^{\left( \mu \nu \right) }=\left( A^{\mu \nu }+A^{\nu \mu }\right) /2$.
Therefore, we can uniquely identify the fundamental fluid dynamical
quantities such as the particle density, energy density and isotropic
pressure: 
\begin{eqnarray}
n &\equiv &u_{\mu }N^{\mu }=\int dKE_{k}\tilde{f}_{k}, \\
e &\equiv &u_{\mu }u_{\nu }T^{\mu \nu }=\int dKE_{k}^{2}\tilde{f}_{k}, \\
p &\equiv &-\frac{1}{3}\Delta _{\mu \nu }T^{\mu \nu }=-\frac{1}{3}\int
dK\left( \Delta ^{\alpha \beta }k_{\alpha }k_{\beta }\right) \tilde{f}_{k}.
\end{eqnarray}%
Similarly the particle diffusion current, the energy-momentum current and
the stress tensor are given by 
\begin{align}
V^{\mu }& \equiv \Delta _{\alpha }^{\mu }N^{\alpha }=\int dKk^{\left\langle
\mu \right\rangle }\tilde{f}_{k},  \label{def_charge_diffusion} \\
W^{\mu }& \equiv \Delta _{\alpha }^{\mu }u_{\beta }T^{\alpha \beta }=\int
dKE_{k}k^{\left\langle \mu \right\rangle }\tilde{f}_{k},
\label{def_momentum_diffusion} \\
\pi ^{\mu \nu }& \equiv T^{\left\langle \mu \nu \right\rangle }=\int
dKk^{\left\langle \mu \right. }k^{\left. \nu \right\rangle }\tilde{f}_{k}.
\label{def_stress_tensor}
\end{align}%
These macroscopic fields, expressed above as momentum integrals over the
non-equilibrium distribution function, $\tilde{f}_{k}$, are part of the
particle four-current and energy momentum tensor of a non-equilibrated
fluid. Some of these quantities are related to their equilibrium
thermodynamical pendants through the so called matching conditions. It is
standard practice to assume that the particle density and energy density are
unchanged from their equilibrium values,
\begin{equation}
n=n_{0},\ e=e_{0},
\end{equation}%
while the isotropic pressure, $p$, separates into two parts: the
thermodynamical pressure, 
\begin{equation}
p_{0}=-\frac{1}{3}\int dK\left( \Delta ^{\alpha \beta }k_{\alpha }k_{\beta
}\right) \tilde{f}_{0k},  \label{def_equilibrium_pressure}
\end{equation}%
and the bulk viscous pressure, 
\begin{equation}
\Pi =-\frac{1}{3}\int dK\left( \Delta ^{\alpha \beta }k_{\alpha }k_{\beta
}\right) \delta \tilde{f}_{k},  \label{def_bulk_pressure}
\end{equation}%
such that $p=p_{0}+\Pi $. Here we introduced the deviation from equilibrium, 
$\delta \tilde{f}_{k}=\tilde{f}_{k}-\tilde{f}_{0k}$. The matching conditions
require that, $\int dK E_{k} \delta \tilde{f}_{k}=\int dKE_{k}^{2} \delta 
\tilde{f}_{k}=0$, while $V^{\mu }\left( \tilde{f}_{k}\right) \equiv V^{\mu
}\left( \delta \tilde{f}_{k}\right) $, $W^{\mu }\left( \tilde{f}_{k}\right)
\equiv W^{\mu }\left( \delta \tilde{f}_{k}\right) $ and $\pi ^{\mu \nu
}\left( \tilde{f}_{k}\right) \equiv \pi ^{\mu \nu }\left( \delta \tilde{f}%
_{k}\right) $. Now, it is clear that only the non-equilibrium deviations lead to
dissipation, since the moments of the equilibrium
distribution function lead to vanishing dissipative quantities $%
\Pi \left( \tilde{f}_{0k}\right) =V^{\mu }\left( \tilde{f}_{0k}\right)
=W^{\mu }\left( \tilde{f}_{0k}\right) =\pi ^{\mu \nu }\left( \tilde{f}%
_{0k}\right) =0$.

For completeness, here we calculate the entropy four-current from Eq. (\ref%
{kinetic_S_mu}) up to first order in deviations from equilibrium $\delta 
\tilde{f}_{k}$, by expanding the entropy four-current around $\tilde{f}_{0k}$.
This expression is given as, 
\begin{eqnarray*}
S^{\mu }\left( \tilde{f}_{k}\right) &=&S_{0}^{\mu }-\int dKk^{\mu }\left(
\alpha _{0}-\beta _{0}k^{\mu }u_{\mu }\right) \delta \tilde{f}_{k}+O(\delta 
\tilde{f}_{k}^{2}) \\
&=&S_{0}^{\mu }-\alpha _{0}V^{\mu }+\beta _{0}W^{\mu }+O(\delta \tilde{f}%
_{k}^{2}),
\end{eqnarray*}%
where $S_{0}^{\mu }=-\alpha _{0}N_{0}^{\mu }+\beta _{0}T_{0}^{\mu \nu
}u_{\nu }+\beta _{0}p_{0}u^{\mu }$ and we made use of the definitions from
Eqs. \ref{def_charge_diffusion} and \ref{def_momentum_diffusion}. Note that
the higher order deviations from $q$-equilibrium can only be calculated once 
$\delta \tilde{f}_{k}$ is specified. The first order result is independent
of such details.

We recall that Eckart \cite{Eckart:1940te} was the first to discuss
relativistic dissipative fluids in a proper way. The energy-momentum tensor
decomposition proposed by him contained no bulk viscous pressure, $\Pi=0$,
and no particle diffusion current, $V^{\mu }=0$, hence the fluid dynamical
flow velocity was fixed to the conserved particles, $u^{\mu }=N^{\mu }/\sqrt{%
N^{\mu }N_{\mu }}$. Moreover the energy-momentum current in this case was
identified with the heat-flow, $q^{\mu }=W^{\mu }$. Later Landau and
Lifshitz \cite{Landau_book} introduced a different decomposition where the
flow is tied to the flow of energy-momentum $u^{\mu }=T^{\mu \nu }u_{\nu }/%
\sqrt{T^{\mu \alpha }u_{\alpha }T_{\mu \beta }u^{\beta }}$. This defines the
flow as the only time-like eigenvector of the energy-momentum tensor.
Therefore the energy-momentum diffusion current vanishes, $W^{\mu }=0$, in
this frame. These physically different choices are related by the general
expression for the heat-flow $q^{\mu }\equiv W^{\mu}-h_{0}V^{\mu }$.

Here we showed that using the NEBE we obtain formally the same results as
in the case of the classical Boltzmann equation, but they are generalized to
include the non-extensivity of entropy.

\section{Transport coefficients in the relaxation time approximation}

Since the equations of dissipative fluid dynamics are not closed due to the
fact that $\tilde{f}_{k}$ is unknown, we need to find additional equations
and relations which close the system of equations. Let us assume, following
Chapman and Enskog \cite{Chapman_book}, that the non-equilibrium
contributions from the streaming term are negligible near the equilibrium $\delta 
\tilde{f}_{k}\ll $ $\tilde{f}_{0k}$, that is $%
\ k^{\mu }\partial _{\mu }\delta \tilde{f}_{k}\rightarrow 0$. 
In this way $k^{\mu }\partial _{\mu }%
\tilde{f}_{k}\simeq $ $k^{\mu }\partial _{\mu }\tilde{f}_{0k}$ and therefore
all non-equilibrium moments vanish from the left hand side of the transport
equation. Hence we obtain the gradients of thermodynamic quantities
multiplied by transport coefficients. 
We calculate the streaming term in $q$-equilibrium, 
\begin{equation}
k^{\mu }\partial _{\mu }\tilde{f}_{0k}=qf_{0k}^{2q-1}\left( k^{\mu }\partial
_{\mu }\alpha _{0}-E_{k}k^{\mu }\partial _{\mu }\beta _{0}-\beta _{0}k^{\nu
}k^{\mu }\partial _{\mu }u_{\nu }\right) .  \label{d_f0}
\end{equation}%

From here we follow the pioneering works of Anderson and Witting 
\cite{Anderson_Witting} and employ the relaxation time approximation. This
simplifies the collision integral of the NEBE, 
\begin{equation}
C\left[ f_{k}\right] \simeq -E_{k}\frac{\left( \tilde{f}_{k}-\tilde{f}%
_{0k}\right) }{\tau _{C}},  \label{Coll_int}
\end{equation}%
where $\tau _{C}$ is the relaxation time. The relaxation time can be
interpreted as a mean time between collisions. Furthermore, this
approximation fixes the local rest frame so that the energy flow vanishes,
as it will be shown below. 
Note that these approximations are made in order to simplify the collision term, 
which in the NEBE is even more complicated than in the classical case. 
Nevertheless, these model equations and solutions can be used to make very reasonable 
first estimates for the transport coefficients. 

Replacing $\tilde{f}_{k}=\tilde{f}_{0k}+\delta 
\tilde{f}_{k}$ into the NEBE and using the previous assumptions we get the
following transport equation, 
\begin{equation}
k^{\mu }\partial _{\mu }\tilde{f}_{0k}=-\frac{E_{k}}{\tau _{C}}\delta \tilde{%
f}_{k}.
\end{equation}%
Utilizing Eqs. (\ref{d_f0}) and (\ref{Coll_int}) we arrive at, 
\begin{eqnarray}
\delta \tilde{f}_{k} &=&\tau _{C}qf_{0k}^{2q-1}\left[ \left( \frac{\beta _{0}%
}{E_{k}}\frac{\theta }{3}\left( k^{\alpha }k^{\beta }\Delta _{\alpha \beta
}\right) -\dot{\alpha}_{0}+E_{k}\dot{\beta}_{0}\right) \right.   \notag \\
&&\left. +\left( h_{0}^{-1}-E_{k}^{-1}\right) k^{\mu }\nabla _{\mu }\alpha
_{0}+\frac{\beta _{0}}{E_{k}}k^{\left\langle \mu \right. }k^{\left. \nu
\right\rangle }\sigma _{\mu \nu }\right] .  \label{delta_f}
\end{eqnarray}%
Here we replaced $\partial _{\mu }u_{\nu }=u_{\mu }\dot{u}_{\nu }+\frac{1}{3}%
\theta \Delta _{\mu \nu }+\sigma _{\mu \nu }+\omega _{\mu \nu }$ where $%
\theta =\nabla _{\mu }u^{\mu }$ is the expansion scalar, $\sigma ^{\mu \nu
}=\nabla ^{\left\langle \mu \right. }u^{\left. \nu \right\rangle }$ shear
stress tensor and $\omega ^{\mu \nu }=\left( \nabla ^{\mu }u^{\nu }-\nabla
^{\nu }u^{\mu }\right) /2$ vorticity. In the last term $\beta
_{0}k^{\left\langle \mu \right. }k^{\left. \nu \right\rangle }\omega _{\mu
\nu }=0$, due to the fact that the vorticity is antisymmetric while$\
k^{\left\langle \mu \right. }k^{\left. \nu \right\rangle }$ is symmetric.
Note that we made use of the conservation laws of perfect fluids to
calculate the proper time derivatives and hence we express them in terms of
gradients. Thus, applying Eq. (\ref{d_J_nk}) for different $i,j$ values we
obtain 
\begin{eqnarray}
\dot{\alpha}_{0} &=&n_{0}\mathcal{D}_{q\left( 2,0\right) }^{-1}\left[ h_{0}%
\mathcal{K}_{q\left( 2,0\right) }-\mathcal{K}_{q\left( 3,0\right) }\right]
\theta ,  \label{alpha_dot} \\
\dot{\beta}_{0} &=&n_{0}\mathcal{D}_{q\left( 2,0\right) }^{-1}\left[ h_{0}%
\mathcal{K}_{q\left( 1,0\right) }-\mathcal{K}_{q\left( 2,0\right) }\right]
\theta ,  \label{beta_dot} \\
\dot{u}^{\mu } &=&\beta _{0}^{-1}\left[ h_{0}^{-1}\nabla ^{\mu }\alpha
_{0}-\nabla ^{\mu }\beta _{0}\right] .  \label{beta_grad}
\end{eqnarray}

Knowing the non-equilibrium distribution function, we can calculate now
dissipative quantities. The stress tensor can be obtained in a
straightforward manner from Eq. (\ref{def_stress_tensor}), together with the
definitions of the integral from Eq. (\ref{K_i_j}), as being 
\begin{eqnarray}
\pi ^{\mu \nu } &=&\tau _{C}\beta _{0}\sigma ^{\alpha \beta }\left( q\int
dKf_{0k}^{2q-1}E_{k}^{-1}k^{\left\langle \mu \right. }k^{\left. \nu
\right\rangle }k_{\left\langle \alpha \right. }k_{\left. \beta \right\rangle
}\right) \notag \\
&=&2\left( \tau _{C}\beta _{0}\mathcal{K}_{q\left( 3,2\right) }\right)
\sigma ^{\mu \nu },
\end{eqnarray}%
whence the traditional Newton-Navier-Stokes relation between the stress
tensor and shear tensor emerges: 
\begin{equation}
\pi ^{\mu \nu }=2\eta \sigma ^{\mu \nu }.
\end{equation}%
The shear viscosity coefficient is given by 
\begin{equation}
\eta =\tau _{C}\beta _{0}\mathcal{K}_{q\left( 3,2\right) }.
\label{shear_visc_coeff}
\end{equation}%
The relativistic generalization of the Newtonian shear $\sigma ^{\mu \nu }$,
and the Navier-Stokes relation between the stress and shear have the same
form for the $q$-modified statistics as the ones obtained using the
classical BG statistics.

Similarly to the shear viscosity the heat-conductivity coefficient can be
calculated. One can show that $W^{\mu
}\sim \left( \mathcal{K}_{q\left( 2,1\right) }-h_{0}\mathcal{K}_{q\left(
3,1\right) }\right) =0$ in the Anderson-Witting relaxation time-approximation. 
Using Eq. (\ref{def_charge_diffusion}) and$\ q^{\mu }=-h_{0}V^{\mu
}$ the result is
\begin{align}
q^{\mu }& =\tau _{C}\left( \nabla ^{\mu }\alpha _{0}\right) \left[
h_{0}\left( \frac{q}{3}\int dKf_{0k}^{2q-1}E_{k}^{-1}\left( k^{\mu }k^{\nu
}\Delta _{\mu \nu }\right) \right) \right.   \notag \\
& \left. -\left( \frac{q}{3}\int dKf_{0k}^{2q-1}\left( k^{\mu }k^{\nu
}\Delta _{\mu \nu }\right) \right) \right]   \notag \\
& =\tau _{C}\left( \nabla ^{\mu }\alpha _{0}\right) \left[ h_{0}\mathcal{K}%
_{q\left( 1,1\right) }-\mathcal{K}_{q\left( 2,1\right) }\right] .
\end{align}%
The Fourier-Navier-Stokes law becomes 
\begin{equation}
q^{\mu }=-\kappa \left( h_{0}^{-1}\beta _{0}^{-2}\nabla ^{\mu }\alpha
_{0}\right) ,
\end{equation}%
where $\nabla ^{\mu }\alpha _{0}=-h_{0}T^{-2}\left( \nabla ^{\mu }T-T\dot{u}%
^{\mu }\right) $ and the coefficient for heat-conductivity or thermal
conductivity is given by 
\begin{equation}
\kappa =\tau _{C}h_{0}\beta _{0}^{2}\left( \mathcal{K}_{q\left( 2,1\right)
}-h_{0}\mathcal{K}_{q\left( 1,1\right) }\right) .  \label{heat_cond_coeff}
\end{equation}%

Finally the bulk viscous pressure also known as the volume or second viscosity, 
can be calculated from Eq. (\ref{def_bulk_pressure}) where we used 
Eqs. (\ref{alpha_dot},\ref{beta_dot}) to express the proper time derivatives 
in Eq. (\ref{delta_f}), hence 
\begin{align}
\Pi & =-\tau _{C}\frac{\theta }{3}\left[ \beta_{0} \frac{q}{3}\int
dKf_{0k}^{2q-1}E_{k}^{-1}\left( \Delta ^{\alpha \beta }k_{\alpha }k_{\beta
}\right) ^{2}\right.  \notag \\
& \left. +n_{0}\left( \frac{\mathcal{K}_{30}-h_{0}\mathcal{K}_{20}}{\mathcal{%
D}_{20}}\right) q\int dKf_{0k}^{2q-1}\left( \Delta ^{\alpha \beta }k_{\alpha
}k_{\beta }\right) \right.  \notag \\
& \left. -n_{0}\left( \frac{\mathcal{K}_{20}-h_{0}\mathcal{K}_{10}}{\mathcal{%
D}_{20}}\right) q\int dKf_{0k}^{2q-1}E_{k}\left( \Delta ^{\alpha \beta
}k_{\alpha }k_{\beta }\right) \right] \notag \\
& =-\tau _{C}\frac{\theta }{3}\left[ 5\beta _{0}\mathcal{K}_{q\left(
3,2\right) }+3n_{0}\left( \frac{\mathcal{K}_{30}-h_{0}\mathcal{K}_{20}}{%
\mathcal{D}_{20}}\right) \mathcal{K}_{q\left( 2,1\right) }\right.  \notag \\
& \left. -3n_{0}\left( \frac{\mathcal{K}_{20}-h_{0}\mathcal{K}_{10}}{%
\mathcal{D}_{20}}\right) \mathcal{K}_{q\left( 3,1\right) }\right] .
\end{align}%
From this the relativistic version of the classical Stokes result follows
immediately: 
\begin{equation}
\Pi =-\zeta \theta ,
\end{equation}%
with the bulk viscosity coefficient,
\begin{eqnarray}
\zeta &=&\tau _{C}\frac{n_{0}}{3\mathcal{D}_{20}}\left[ 5\frac{\beta _{0}}{n_{0}}%
\mathcal{D}_{20}\mathcal{K}_{q\left( 3,2\right) }+3\left( \mathcal{K}%
_{30}-h_{0}\mathcal{K}_{20}\right) \mathcal{K}_{q\left( 2,1\right) }\right. 
\notag \\
&&\left. -3\left( \mathcal{K}_{20}-h_{0}\mathcal{K}_{10}\right) \mathcal{K}%
_{q\left( 3,1\right) }\right] .  \label{bulk_visc_coeff}
\end{eqnarray}%
Once again we note that the above results formally resemble those of the BG
statistics, the only explicit $q$-dependence occurs in the definitions of
the $q$-modified thermodynamic integrals. All formulas show that the
transport coefficients are directly proportional to the mean free time
between collisions and so inversely proportional to the cross section.

Having defined the dissipative quantities, the deviation from the $q$%
-equilibrium Eq. \ref{delta_f} can be written as,
\begin{equation}
\delta \tilde{f}_{k}=qf_{0k}^{2q-1}\left( C_{\Pi }\Pi +C_{q}k^{\mu }q_{\mu
}+C_{\pi }k^{\mu }k^{\nu }\pi _{\mu \nu }\right) ,  \notag
\end{equation}%
where $C_{\Pi }$, $C_{q}$ and $C_{\pi }$ are tedious expression inversely 
proportional to the respective transport coefficients, and may be calculated 
from the above formulas.

\section{Non-extensive effects in the transport coefficients}

In this section we investigate the previously obtained results in the LRF where $%
E_{k}|_{LRF}=\sqrt{m^{2}+k^{2}}$. The thermodynamic
integrals are calculated in hyperspherical coordinates, $k^{\mu
}k_{\mu }=m^{2}$, $k^{\mu }u_{\mu }\equiv k^{0}=m\cosh \chi $, $%
k^{2}=m^{2}\sinh ^{2}\chi $, $\Delta ^{\mu \nu }k_{\mu }k_{\nu
}=-m^{2}\sinh ^{2}\chi $ and $dK=\frac{g}{\left( 2\pi \right) ^{3}}4\pi
m^{2}\sinh ^{2}\chi d\chi $. Note that the integrals for $q\geq 1$ are
calculated up to infinity while for $q<1$ the upper limit of integration is $%
\chi _{\max }=\cosh ^{-1}\left( \frac{T}{m\left( 1-q\right) }\right) $.
Furthermore, the chemical potential is set to zero $\mu =0$, while the spin
degeneracy is $g=1$.
\begin{figure}[hbt!]
\centering
\includegraphics[width=8.cm]{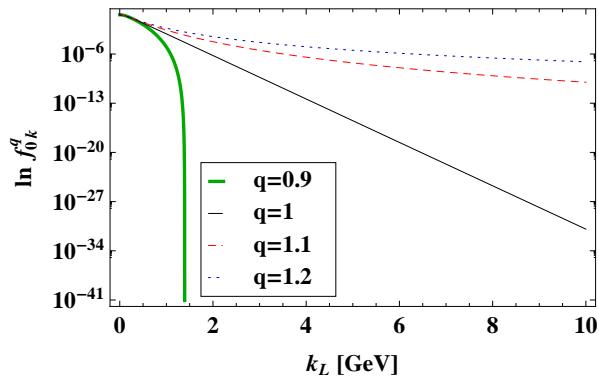} 
\caption{(Color online) The $\tilde{f}_{0k}$ distribution as function of longitudinal
momenta for different $q$-parameters: $q=0.9$ (thick line), $q=1$ (thin
line), $q=1.1$ (dashed line) and $q=1.2$ (dotted line). The mass of
particles as well as the temperature are $m=T=140$ MeV.}
\label{fig:fq_Ek}
\end{figure}

In Fig. \ref{fig:fq_Ek} we plot the natural logarithm of the $q$-th power of the $q$-equilibrium 
distribution function from Eq. (\ref{q_equilibrium}) in the LRF, with $m=140$ MeV and $T=140$ MeV
for longitudinal momenta $k_{L}\in \left[ 0,10\right] $ GeV and 
transverse momenta $k_{T}=0$. The functions are plotted for $%
q=0.9$ (thick line), $q=1$ (thin line), $q=1.1$ (dashed line) and $q=1.2$
(dotted line). Note that for $q=0.9$ the function is practically cut-off at $%
k_{L}=m\sinh \chi _{\max }$.

\begin{figure}[thb!]
\centering
\includegraphics[width=8.cm]{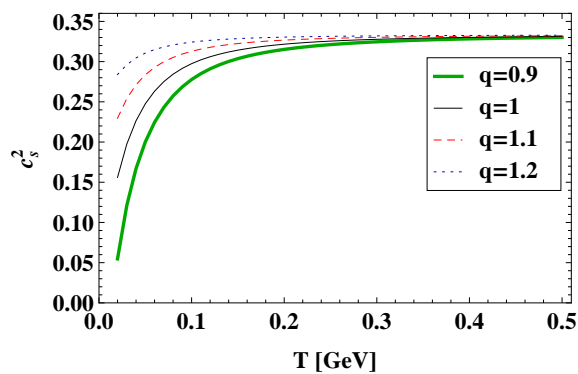} 
\caption{(Color online) The $c^2_s$ as a function of temperature for different $q$%
-parameters: $q=0.9$ (thick line), $q=1$ (thin line), $q=1.1$ (dashed line)
and $q=1.2$ (dotted line). }
\label{fig:cs2}
\end{figure}
\begin{figure}[hbt!]
\centering
\includegraphics[width=8.cm]{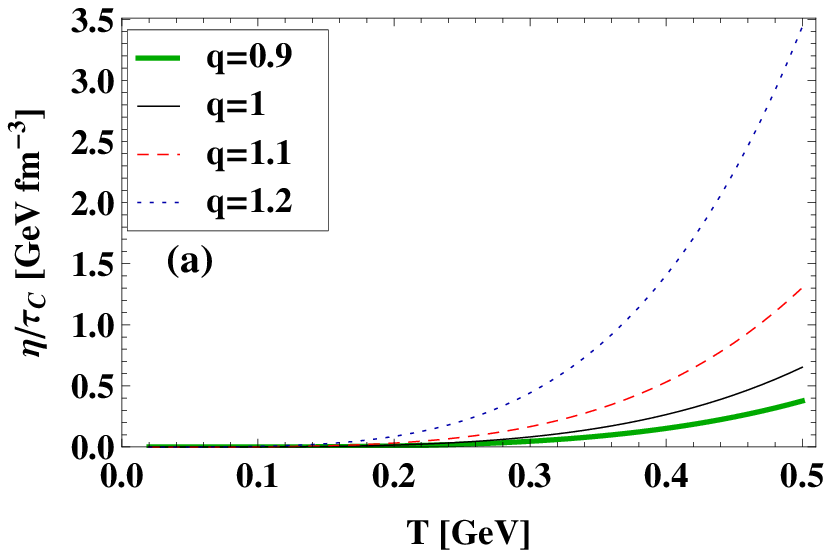} %
\includegraphics[width=8.cm]{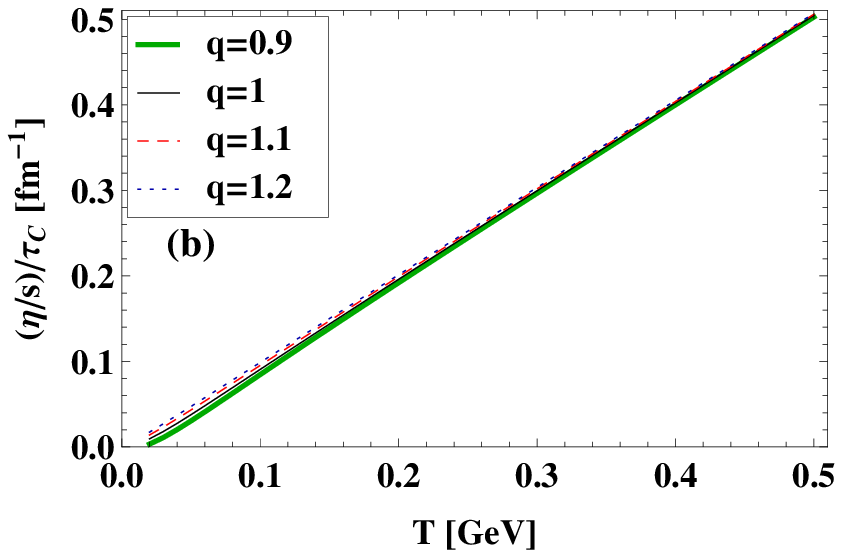} 
\caption{(Color online) The shear viscosity coefficient divided by the relaxation time, $%
\protect\eta \protect\tau _{C}^{-1}$ (a) and the shear viscosity coefficient
over entropy density divided by the relaxation time $\left(\protect\eta/s
\right) \protect\tau _{C}^{-1}$ (b) as function of temperature for different 
$q$-values.}
\label{fig:eta}
\end{figure}

Next we plot the speed of sound squared $c_{s}^{2}$ as a function of
temperature and $q$-parameter from Eq. (\ref{cs_2}). The mass of particles
is fixed at $m=140$ MeV, while the temperature is $T\in \left[ 2,500\right] $
MeV.
We see that for any value of $q$ the speed of sound is increasing with
temperature while it saturates for $T \gg m$ to the ultra-relativistic limit, 
$c_{s}^{2}=1/3$ independently of $q$. This is a straightforward result from
Eq. (\ref{cs_2}). The speed of sound is larger for larger $q$ at any given
temperature.

\begin{figure}[thb!]
\centering
\includegraphics[width=8.cm]{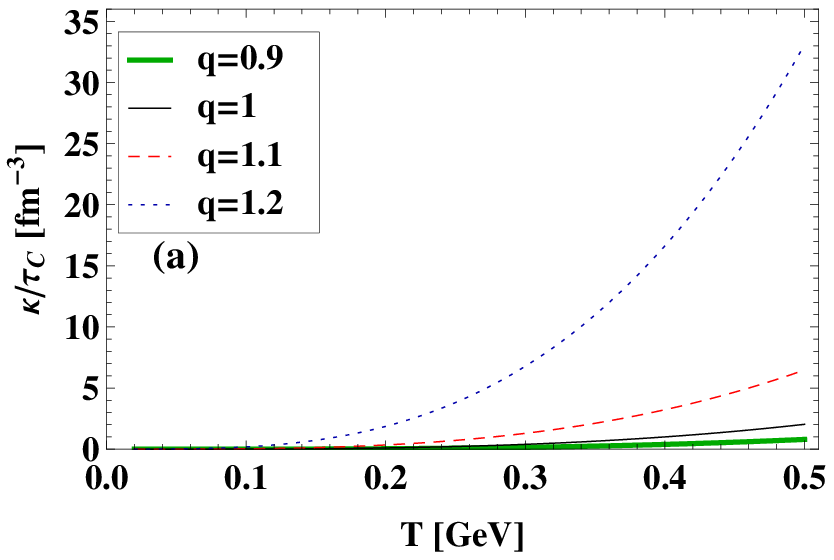} 
\includegraphics[width=8.cm]{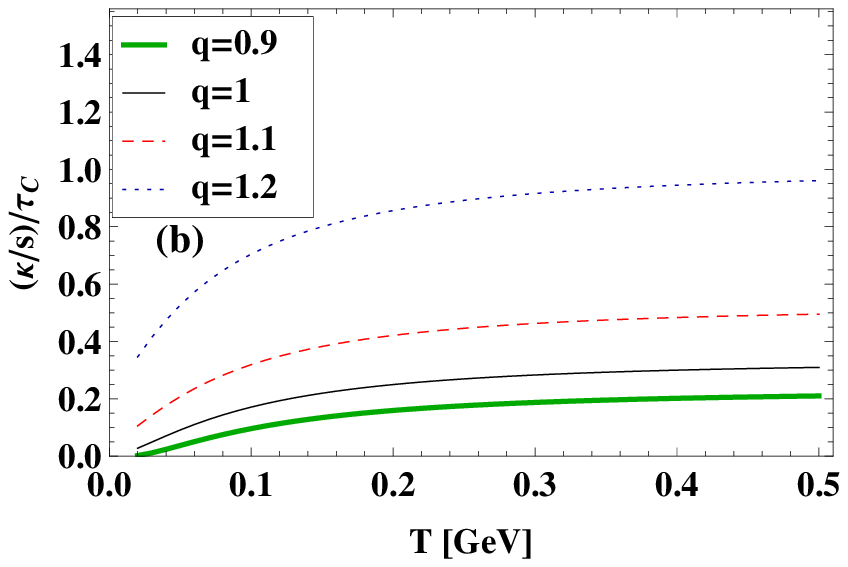} 
\caption{(Color online) The heat-conductivity coefficient divided by the relaxation time
(a) and the heat-conductivity coefficient to entropy density ratio divided
by the relaxation time as a function of temperature for different $q$%
-values. }
\label{fig:kappa}
\end{figure}

The shear viscosity divided by the relaxation time from Eq. (%
\ref{shear_visc_coeff}), $\eta \tau _{C}^{-1}=\beta _{0}\mathcal{K}%
_{q\left( 3,2\right) }$, is shown in Fig. \ref{fig:eta}a. On the other
hand the shear viscosity to entropy ratio divided by the relaxation time $%
\frac{\eta }{s}\tau _{C}^{-1}=\frac{\mathcal{K}_{q\left( 3,2\right) }}{%
\mathcal{J}_{q\left( 2,0\right) }-\mathcal{J}_{q\left( 2,1\right) }}$ can be
inspected for different values of $q$ in Fig. \ref{fig:eta}b.

We observe that the viscosity increases with increasing temperature for all $%
q$-values. This behaviour is more pronounced for larger $q$-values, hence
there is a monotonic increase in the viscosity coefficient with increasing $%
q $ at a given temperature. At the same time it turns out that the shear
viscosity to entropy ratio at a given temperature is almost insensitive to
different values of $q$. Fig. \ref{fig:eta}b demonstrates that for a given
relaxation time $\tau _{C}$ the $\eta /s$ ratio does not change
significantly and at high temperatures the $q$-dependence diminishes. 
This also means that the ratio of dissipative to equilibrium quantities such as 
$\pi ^{\mu \nu }/\left( e_{0}+p_{0}\right) \simeq \left( \eta /s\right) /T$
does not change at high temperatures. This is due to the
fact that the entropy density also increases with increasing $q$-values by
about the same factor. In the classical Maxwell-Boltzmann massless limit $%
\left( \eta \tau _{C}^{-1}\right) _{q=1}=\beta _{0}I_{\left( 3,2\right)
}=4/\left( 15e_{0}\right) $ and $s_{\left( q=1\right) }=\beta _{0}\left(
I_{\left( 2,0\right) }-I_{\left( 2,1\right) }\right) =4\beta _{0}e_{0}/3$ we
recover the familiar results \cite{Gavin:1985ph}.

Similarly to the viscosity coefficient, we also plot the coefficient for the
heat-conductivity from Eq. (\ref{heat_cond_coeff}), as well as the ratio of
heat-conductivity to entropy density divided by the relaxation time. Both of
these quantities, shown in Figs. \ref{fig:kappa}a and \ref{fig:kappa}b,
are sensitive to the value of the parameter $q$.

Furthermore, we observe that the heat-conductivity increases with
increasing $q$-values, and - other than before - the ratio $\frac{\kappa }{%
s}\tau _{C}^{-1}$ increases in a noticeable fashion.

\begin{figure}[thb!]
\centering
\includegraphics[width=8.cm]{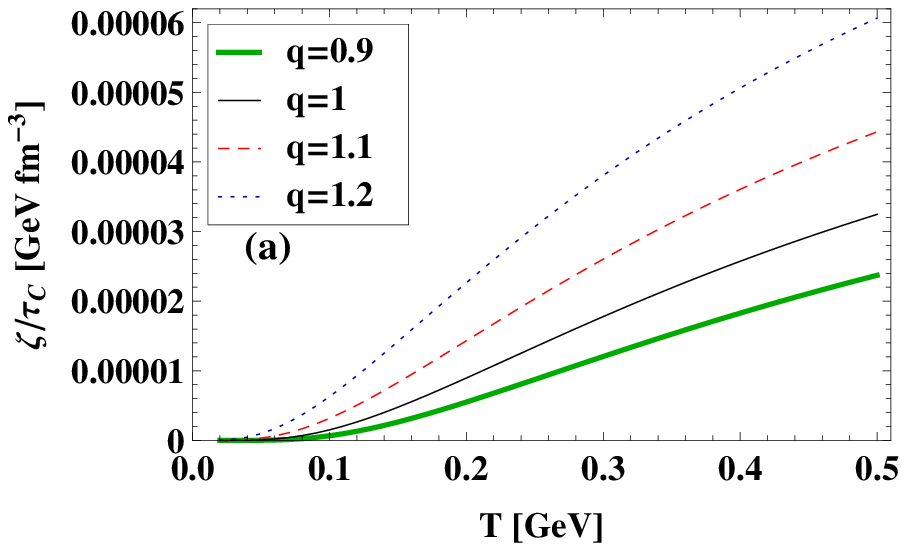} 
\includegraphics[width=8.cm]{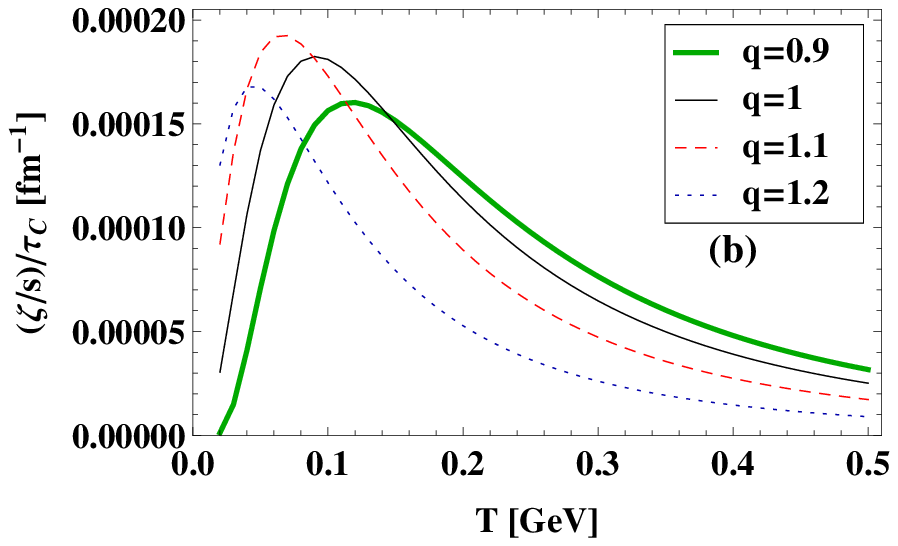} 
\caption{(Color online) The bulk viscosity coefficient divided by the relaxation time (a)
and the bulk viscosity coefficient to entropy density ratio divided by the
relaxation time as a function of temperature for different $q$-values.}
\label{fig:zeta}
\end{figure}

The last figures, Figs. \ref{fig:zeta}a and Fig. \ref{fig:zeta}b show the
bulk viscosity coefficient divided by the relaxation time, and the bulk
viscosity coefficient to entropy density ratio divided by the relaxation
time.

We observe that the value of bulk viscosity coefficient is very small. It
is about $10^{-3}$ times smaller than the coefficient of shear viscosity 
$\zeta \simeq 10^{-3}\eta $, which corroborates the classical result that 
the bulk viscosity in the non-relativistic or ultra-relativistic
massless limit vanishes, while it is much smaller than the shear viscosity
in between for massive particles \cite{Gavin:1985ph,Israel_1963}. 
On the other hand the bulk viscosity to entropy density ratio does show 
a monotonic increase with increasing $q$ but only for $T \ll m$. 
For $T > m$ the behaviour is reversed due to the fact that 
the entropy density increases with temperature at a slower rate compared to the viscosity.

\section{Conclusions and outlook}

In this paper we have calculated the transport coefficients from a $q$%
-modified Boltzmann equation. The results show that all transport
coefficients increase monotonically with increasing $q$-values for relativistic 
ideal gases. 
The speed of sound is also increasing with increasing $q$ but it saturates quickly to the
well known ultra-relativistic limit independent of $q$.

Furthermore, we derived the $q$-generalized versions of the classical
Navier-Stokes-Fourier equations of relativistic dissipative fluid dynamics from a $q$%
-generalized Boltzmann transport equation. These equations relate the
dissipative quantities to the thermodynamical forces linearly with positive
transport coefficients which dependent on parameter $q$.

The resulting equations, are nevertheless parabolic, hence acausal and
unstable \cite{his}. 
Hyperbolicity related problems are not solved by introducing the parameter $q$.
This manqu\'e can be resolved by an appropriate choice of the relaxation time 
in the evolving fluid \cite{Denicol:2011fa}.
Superior but more complicated equations of motion may be derived from the NEBE 
as in the classical theory, with methods different from the one presented in this work, 
see for example Refs. \cite{Grad,Israel:1979wp,Romatschke:2009im,Denicol:2010xn,Betz:2010cx}.
\\

\section*{Acknowledgments}

The authors thank for G.S. Denicol, H. Niemi, L.P. Csernai, M. Gyulassy for 
discussions and P. V\'an for reading the manuscript and for valuable comments. 
This work was supported by the Hungarian National Scientific Fund OTKA K68108. 
E.M. was supported by the National Development Agency 
OTKA/NF\"U 81655.



\end{document}